\documentclass[twocolumn]{aastex63}

\accepted{September 19, 2019}
\submitjournal{ApJ}

\shortauthors{M. De Furio et al.}
\shorttitle{M-type Binaries in the ONC}
\graphicspath{{./}{figures/}}

\begin{document}

\title{A Search for Intermediate Separation Low Mass Binaries in the Orion Nebula Cluster}

\correspondingauthor{Matthew De Furio}
\email{defurio@umich.edu}

\author{Matthew De Furio}
\affiliation{Department of Astronomy, University of Michigan, Ann Arbor, MI}

\author{Megan Reiter}
\affiliation{UK Astronomy Technology Centre, Edinburgh, United Kingdom}

\author{Michael R. Meyer}
\affiliation{Department of Astronomy, University of Michigan, Ann Arbor, MI}

\author{Alexandra Greenbaum}
\affiliation{Department of Astronomy, University of Michigan, Ann Arbor, MI}

\author{Trent Dupuy}
\affiliation{Gemini Observatory, Hilo, HI}

\author{Adam Kraus}
\affiliation{Department of Astronomy, University of Texas - Austin, Austin, TX}



\begin{abstract}

We present the results of a binary population study in the Orion Nebula Cluster (ONC) using archival Hubble Space Telescope (HST) data obtained with the Advanced Camera for Surveys (ACS) in Johnson V filter (HST Proposal 10246, PI M. Robberto).  Young clusters and associations hold clues to the origin and properties of multiple star systems.  Binaries with separations $< 100 $ AU  are useful as tracers of the initial binary population since they are not as likely to be destroyed through dynamical interactions.  Low mass, low stellar density star-forming regions such as Taurus-Auriga, reveal an excess of multiples compared to the Galactic Field.  Studying the binary population of higher mass, higher stellar density star-forming regions like the ONC provides useful information concerning the origin of the Galactic Field star population.  In this survey, we characterize the previously unexplored (and incomplete) separation parameter space of binaries in the ONC (15 - 160 AU) by fitting a double-PSF model built from empirical PSFs.  We identified 14 candidate binaries (11 new detections) and find that 8$_{-2\%}^{+4\%}$ of our observed sample are in binary systems, complete over mass ratios and separations of 0.6 $< $ q $< $ 1.0 and 30 $< $ a $< $ 160 AU.  This is consistent with the Galactic Field M-dwarf population over the same parameter ranges, 6.5\% $\pm$ 3\%.  Therefore, high mass star forming regions like the ONC would not require further dynamical evolution for their binary population to resemble the Galactic Field, as some models have hypothesized for young clusters.

\end{abstract}



\section{Introduction} \label{sec:intro}

Multiplicity is a common outcome of star formation that impacts stellar evolution, planet formation, and the eventual makeup of the Galactic Field population.  Binary systems form when the stars themselves form, through disk fragmentation \citep{Adams1989,Bonnell1994} and turbulent fragmentation \citep{Goodwin2004,Offner2010}.  By studying young, star-forming regions, we can identify the properties of primordial binary populations.  These populations can help us put constraints on how binaries form as well as how they evolve with time due to processes like dynamical interactions.  Stars tend to form in clusters or associations and by studying diverse star forming regions, we can search for differences in binary populations as a function of various environmental properties such as cluster size and stellar density.

Many surveys have been carried out in the nearest young associations in order to probe their binary populations.  Surveys of Taurus-Auriga, Ophiuchus-Scorpius, Chameleon, Lupus, and Corona Australis all find evidence for an excess in binary frequency of these star forming regions when compared to the Galactic Field population \citep{Ghez1993,Leinert1993,Simon1995,Ghez1997,Brandner1996,ReipurthZinnecker1993}.  In one of the most complete binary surveys of a young association, \citet{Kraus2011} found that Taurus-Auriga has an excess of companions around low mass stars with separations 3 - 5000 AU compared to the Galactic Field population.

These binary excesses have led to the speculation that many stars could form as multiple systems where a large fraction of these binaries could be destroyed through dynamical evolution \citep{Horton2001,kroupa1995}.  Importantly, all of these young associations have low stellar densities \citep[ $\sim$ 10$^{1.5}$ - 10$^{3}$ pc$^{-3}$, see][]{Liu2003} which could allow for an excess in wide binaries that are not destroyed through dynamical evolution.  Previous analyses of young star forming regions provide evidence for a larger number of wide binaries in lower stellar density regions compared to higher stellar density regions \citep{Simon1997,KrausHill2008}.  As suggested by \citet{patience2002}, low stellar density regions could contribute more wide binaries to the Galactic Field population than higher stellar density regions like the Orion Nebula Cluster (ONC), \citep[ $\sim 10^{4.5}$ pc$^{-3}$, see][]{Liu2003}, which could contribute more binaries with separations $\lesssim$ 100s AU, while even higher density regions (e.g. Westerlund 1, $\sim 10^{5}$ pc$^{-3}$) could contribute more of the tight binaries.  The Galactic Field binary properties would then be an average with respect to the contribution of each type of star formation environment.

The ONC is the nearest high mass star forming region at $\sim$ 400 pc \citep{oriondistance}, and thus the ideal location to study the primordial binary population over a broad range of separations in a more dense environment than the nearby young associations.  It has also been extensively observed through the Hubble Space Telescope Treasury Program \citep{Robberto2013} using the Advanced Camera for Surveys (ACS), granting us access to a large survey area with high angular resolution (a diffraction limit of $\sim$ 0.06").

Previous surveys have attempted to characterize the binary population of the ONC.  A direct imaging survey \citep{Reipurth2007} using ACS/WFC on HST was sensitive from 67.5 - 675 AU, but used the H$_{\alpha}$ narrow band filter, making it difficult to estimate masses.  Other surveys, like \citet{Kounkel2016_100}, searched for companions around protostars and pre-main-sequence stars and were sensitive to companions between 100 - 1000 AU.  A radial velocity survey by \citet{Kounkel2016_Spec} searched for companions around primary stars with temperatures $\geq$ 4000 K and were $\sim$ 20\% complete out to 4 AU and 5\% complete out to 10 AU.

An adaptive optics (AO) survey of the ONC by \citet{Kohler2006} achieved minimum separations of 60 AU and was sensitive to mass ratios down to 0.4 for separations 160-500 AU.  The most recent survey of the ONC was carried out by \citet{Duchene2018} with AO and is sensitive over separations 10 - 60 AU.  Their sample of 42 sources mostly consisted of solar-type stars, and had 11 M-stars, masses $< $ 0.6 \(\textup{M}_\odot\).  Low mass stars ($< $ 0.6 \(\textup{M}_\odot\)) dominate the stellar population of the ONC in both total mass and number, and therefore play an important role in dynamical evolution.  It follows that a much larger sample of M-stars is necessary to determine if an excess in companion frequency is typical of most star forming regions as seen in Taurus-Auriga and other low density associations which are also dominated by low mass stars.  The HST Treasury Program of the ONC has the sensitivity to observe low mass M-stars and, with PSF-fitting, e.g. \citet{Garcia2015}, identify companions at separations between 15 - 160 AU, a parameter space previously unexplored and incomplete for low mass stars.

In Section 2, we describe the ONC sample, our PSF-fitting model, and the sensitivity of our survey.  In Section 3, we present our discovered binaries as well as the comparison to the Galactic Field population.  Section 4 lays out the comparisons of the ONC binary population to that of Taurus-Auriga and previous ONC surveys in addition to discussing the implications of this survey.  In Section 5, we summarize our conclusions.

\section{Methods} \label{sec:methods}
\subsection{The Data} \label{subsec:data}
We downloaded broadband data from the Treasury Program of the Orion Nebula Cluster (GO program 10246, PI: M. Robberto) from the Hubble Space Telescope (HST) archive.  Our analysis utilizes the data taken in the F555W filter (Johnson V band) with the Advanced Camera for Surveys in the Wide Field Channel mode (ACS/WFC) over a series of 104 orbits during Cycle 13.  ACS has a plate scale of 0.05"/pixel.  The Treasury Program survey covered $\sim$ 600 square arcminutes where most of the area appears in at least two separate exposures, each with an integration time of 385s.  From these data, we analyzed cluster members that fit the selection criteria described in Sec. 2.3.  A detailed explanation of observations and data reduction can be found in \citet{Robberto2013}.

\subsection{The Model} \label{subsec:model}
We now describe how we implemented empirical PSF-fitting to potential binaries.  \citet{AndersonKing2000,AndersonKing2003,AndersonKing2004,AndersonKing2006}, hereafter AK06, developed 90 empirically derived PSFs per filter for ACS/WFC, which enable accurate astrometry of $\sim$ 0.5 mas (0.01 pixels).  AK06 produced a PSF library which describes the shape of the PSF as a function of location on the ACS detector.  These PSFs are 4x super-sampled where the 100x100 pixel array of each PSF corresponds to a 25x25 pixel array on the ACS detector.  As described in AK06, when fitting a PSF to the data, the value of each pixel in the fit is a result of a bicubic interpolation of the super-sampled empirical PSF. In conjunction with this interpolation, the final PSF fit is derived from a linear combination of the four nearest PSFs that are weighted based on proximity to the center of the source in question.  Additionally, AK06 include a PSF perturbation function which alters the shape of the empirical-PSF models based on the PSFs of the brightest sources in an individual data image.  This procedure mitigates the effects of changes in focus and instabilities in pointing which can cause the PSF to deviate from the average over time.

We used the interpolation method of AK06 to make a double-PSF model with their position-dependent empirical PSFs to directly detect binary systems down to separations 60\% of the diffraction limit ($\sim$ 0.06" for the F555W filter on HST).  The companion PSF is defined by its relative location from the center of the primary source and the difference in brightness between the two sources.  Our complete binary model has six parameters: x and y position of the center of the primary, combined magnitude of the binary, pixel separation between the centers of the primary and companion PSFs, position angle between the centers of the primary and companion PSFs, and the difference in magnitude between the primary and companion.

Before fitting the model to the data, we subtract off the mean background around the source inside an annulus with an inner radius of 8 pixels and an outer radius of 14 pixels.  See Sec. 2.5 for a discussion of the limitations of our model in a spatially-varying background environment.  To determine a best fit binary model to any input source, we fit the binary model to the background-subtracted 17x17 pixel stamp of the data where our algorithm first uses a coarse grid search to identify a region of parameter space where our model returns the best fit.  Then by defining this result as the initial guess, we use the IDL downhill simplex function, AMOEBA \citep{Press2007}, which incorporates a series of reflections, expansions, and contractions, to minimize the chi squared test statistic.  We calculated the reduced chi squared test statistic of each source by summing the chi squared statistic over each pixel in a 17 x 17 pixel stamp:

\begin{equation}
\chi^{2}_\nu = \frac{1}{\nu-1} \sum\limits_{i=1}^{N} \frac{ (data_{i} - model_{i})^2 }{ \sigma_{i}^{2} }
\end{equation}

Here, the total error per pixel was the summed in quadrature errors of the source photon noise, the standard deviation in the background, the error in the mean of the background, the dark current, and the read noise as seen in Eq. 2.  We added an additional 2\% noise floor, defined as 2\% of the flux in a given pixel, to the source photon noise in order to account for the error in the PSF models.

\begin{equation}
    \sigma^{2} = \sigma_{Source}^{2} + \sigma_{Bkgd}^{2} + \sigma_{\overline{Bkgd}}^{2} + \sigma_{DC}^{2} + \sigma_{RN}^{2}
\end{equation}

\medskip
Our algorithm will always find a "best fit" to any source.  Whether the source in question is truly a binary depends on the sensitivity of our model for the particular S/N of the source.  Our algorithm can converge to a fit that has large differences in magnitude between the primary and potential companion which would correspond to a companion indistinguishable from background noise.  Those binary fits are rejected as they fall outside the sensitivity range, defined in Sec. 2.4, and therefore are classified as singles.

\subsection{The Sample} \label{subsec:sample}
We adopted our input target list from \citet{DaRio2016} based on the Sloan Digital Sky Survey APOGEE INfrared Spectroscopy of Young Nebulous Clusters program (IN-SYNC) survey.  Their high resolution H-band spectroscopic survey of the Orion A molecular cloud includes both potential members and previously known members from the literature detected through various methods, such as infrared excess and X-ray emission, e.g. \citet{Megeath2012,Getman2005}.  Stellar properties for each source were calculated by \citet{DaRio2016} using a fitting procedure thoroughly defined in \citet{Cottaar2014}.  Their procedure estimated the T$_{eff}$, extinction, age, and other stellar parameters.

From this list of ONC members, we made sample cuts for sources that were detectable, below a threshold pixel value, and appeared in the area covered by the HST Treasury Program of the ONC.  To set our pixel value threshold, we fit our model to known binaries in the data, identified by \citet{Reipurth2007}.  Above the peak pixel value of 29,000, the PSF quality became an issue in fitting the binary model to known sources.  Therefore, we set a cutoff at this peak value, giving us 135 sources.  Next, we mandated that the mass of each source (as estimated by Da Rio et al. 2016) be $< $ 0.6 \(\textup{M}_\odot\) in order to define a low mass sample since binary properties (such as projected separation) depend on primary mass, e.g. \citet{Janson2012}, \citet{Raghavan2010}, \citet{DeRosa2014}.  This cut resulted in 113 sources.

After running our double-PSF binary model on all 113 sources, the chi squared values of the total sample followed a chi squared distribution of a six-parameter model.  We set the cutoff chi squared value equal to where the p-value equals 0.1 ($\chi^{2}_{\nu}$ = 1.774).  Below this value, we do not reject the null hypothesis of the data coming from a chi squared distributed sample.  Therefore, the binary fits accurately model the PSF of the sources.  Our last cut resulted in a sample of 101 unique sources, all of which can be described as Pre-Main Sequence (PMS) M-stars.  

There seemed to be four types of sources that were excluded from our sample based on the goodness of fit criteria ($\chi^{2}_{\nu}$ cutoff).  First, potential triples would return a high $\chi^{2}_{\nu}$ as the flux from the tertiary star would make the residuals high.  Next, sources in regions of high nebulosity could also return a high $\chi^{2}_{\nu}$ as the residuals could remain very high.  Also, sources that only appear in one image could return a high $\chi^{2}_{\nu}$ if a cosmic ray hit the detector at the exact location as the source.  This would cause the structure of the perceived PSF to be very different from the model and result in a poor fit.  These three possibilities are discussed further in Sec. 2.5.  Lastly, very high S/N sources were excluded as their PSF structure is less well-defined and result in high $\chi^{2}_{\nu}$.

\begin{figure}[ht!]
\plotone{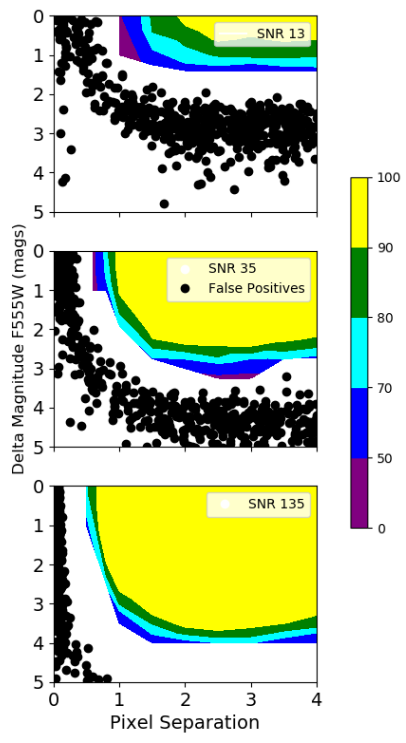}
\caption{Top: Completeness map for sources with S/N = 13. Middle: Completeness map for sources with S/N = 35.  Bottom: Completeness map for sources with S/N = 135.  Each completeness map is color-coded based on the percent of artificial binaries recovered by our algorithm.  False positives are displayed as the black circles in the images. The contrast sensitivity flattens out at further separations as the companion PSF is completely separate from the primary PSF and the sensitivity becomes background limited. \label{fig:Cmap_SN}}
\end{figure}

\subsection{Completeness}\label{subsec:completeness}
In order to distinguish between a real binary and a false positive result from our algorithm, we had to determine the sensitivity of our method.  This is a function of separation and difference in magnitude, as well as S/N of each source.  We created artificial binaries from the PSF models of AK06 based on the location of the source.  To create the companion source, we add another PSF at the defined location and with the specified flux ratio relative to the primary.  To mimic a background environment, we add Poisson noise to each pixel until the artificial binary was at the desired S/N.  

Since the data on ACS/WFC is under-sampled, interpolation of the data to create sub-pixel shifts in the PSF would change the distribution of flux in each pixel and introduce more error.  The PSF models of AK06 are super-sampled, and we can create artificial binaries at sub-pixel separations from the models 
instead of re-sampling the data.  This allows us to accurately replicate the PSF of a binary system without losing information about the PSF if we interpolated the real data.  We created binaries at multiple separations, random position angles, and differences in magnitude depending on the desired S/N of the sample.  

The S/N ratio of each source was calculated with the following equation over all pixels in the 17x17 pixel stamp that encompassed each source where the error is the same as described in Sec. 2.2:

\begin{equation}
    S/N = \frac{\sum\limits_{i=1}^{N}  (data_{i} - \overline{bkgd})}{\sqrt{\sum\limits_{i=1}^{N}\sigma^{2}_{i}}}
\end{equation}

The magnitude bins we sampled were dependent on the S/N.  For all S/N completeness maps, we sampled pixel separations = 1.0 - 4.0 in steps of 0.5 pixels.  For S/N = 21, 35, and 60, we also sampled pixel separations = 0.6 and 0.8 pixels.  For S/N = 90, 113, and 135, we also sampled pixel separations = 0.5 and 0.8 pixels.  For separations 1.5 - 4.0 pixels, the magnitude bin size was generally 0.25 mag.  For pixel separations $\leq$ 1.0 pixel, the magnitude bin size was between 0.4 - 0.75 mags depending on the S/N.

Every binary was made at a random position angle with Poisson noise added until the S/N reached the desired level.  We created 500 binaries in each bin for each S/N and ran the double-PSF code on these sources.  If the output pixel separation was within 0.2 pixels of the known separation and the output difference in magnitude was within 0.2 mags of the known difference in magnitude, then that source would be defined as recovered.  After fitting all the artificial binaries with our double-PSF algorithm, we determined the recovery rates of our method as a function of S/N, separation, and contrast, e.g. Fig. \ref{fig:Cmap_SN}.

For companions at the same pixel separation, it is assumed that the recovery rate of a smaller difference in magnitude companion (e.g. 0 mag) is at least equal to that of a larger difference in magnitude companion (e.g. 3 mag).  Similarly, the recovery rate of larger separation companions (e.g. 4 pixels) can be assumed to be at least equal to that of smaller separation companions (e.g. 1 pixel) for the same difference in magnitude.  We constructed completeness maps based on these recovery rates by interpolating between the bins over the 2D surface of separation and difference in magnitude.  As a check on our method, we created integer-separation binaries constructed from real data and obtained completeness maps consistent with the completeness maps obtained from using the empirical PSF models.

We have created seven completeness maps for S/N = 13, 21, 35, 60, 90, 113, and 135, three of which are shown in Fig.\ref{fig:Cmap_SN}.  To determine the contrast achievable for each source, we assume that the sensitivity of our code on a given source will be at least as good as the nearest completeness map with a lower or equal S/N than the source.  These completeness maps from lowest to high S/N were assigned to 17, 11, 18, 15, 12, 14, and 14 sources (totaling 101) respectively.

We define a recovery rate of 90\% (see Fig. \ref{fig:Cmap_SN}) as the threshold for a candidate binary detection.  This limit allows us to both be confident in the ability of the code to recover the parameters of the binary fit and put a constraint on potential false negative reporting from the code, i.e. $\leq$ 10\% of real binaries that exist in the 90\% recovery rate area would not be detected.  The completeness maps follow the expected trend that our algorithm achieves greater overall contrast for higher S/N primaries as well as resolves companions at smaller separations compared to low S/N primaries.

\subsection{Limitations of the Model}\label{subsec:limitations}
We recognize that there are limits to the application of this code for multiplicity surveys.  The most obvious and important limitation is that our model cannot detect triple or higher order systems.  A true visual triple system would produce high residuals when fit with our model.  Such systems are outside the scope of our study.

An additional complication of our model comes in regions of high, structured background.  The residual background in the data will decrease the S/N and can potentially mask the existence of a companion.  This is especially important in the centers of young, star-forming regions.  Future work refining our model will include fitting a 2D surface to local background in order to mitigate the effects of nebulosity, resulting in better fits for those particular sources in structured background environments.

Lastly, some of the sources in our sample were only found in one frame. It is possible that a cosmic ray could have hit the detector at the same location as the source or the candidate companion.  This effect would result in a poor fit because the observed structure of the PSF would be altered by the cosmic ray and poorly fit with the PSF models.  This is unlikely to have impacted our survey.

\section{Results} \label{sec:results}

\subsection{Detections} \label{subsec:detections}
\begin{figure*}[t]
\plotone{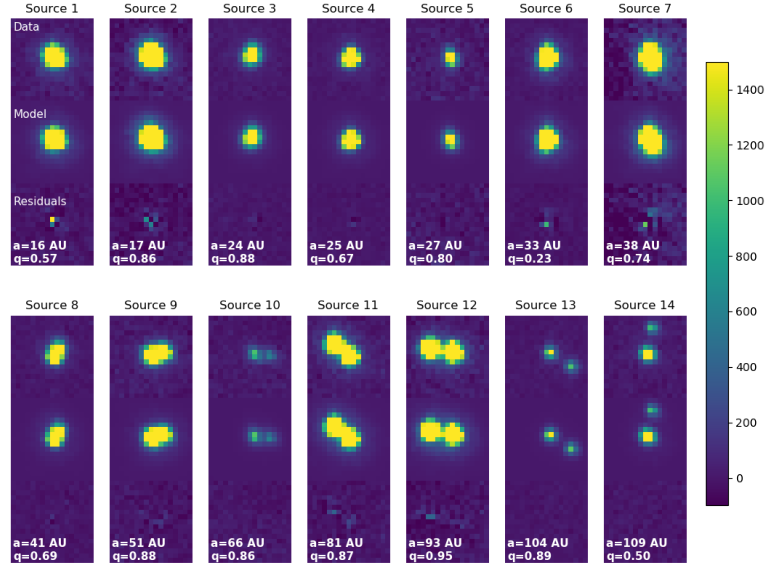}
\caption{Candidate binaries from the sample of Orion members.  In each panel, the top image is the HST data, the middle is the binary fit, and the bottom shows the residual.  Each image has dimensions 17x17 pixels (0.85"x0.85").  The projected separation (a) and mass ratio (q) are displayed on the bottom image of each panel.  Data are in units of Data Number (DN).  The position angle of the image y-axis of Sources \#2, 3, 5, 6, 7, and 10 are -82.2 degrees East of North. The position angle of the image y-axis of Sources \#1, 4, 8, 9, 12, 13, and 14  are 97.8 degrees East of North. The position angle of the image y-axis of Source \#11  are 104.8 degrees East of North.  \label{fig:ex_binaries}}
\end{figure*}

\begin{figure}[ht!]
\includegraphics[height=7.2cm,width=9.0cm]{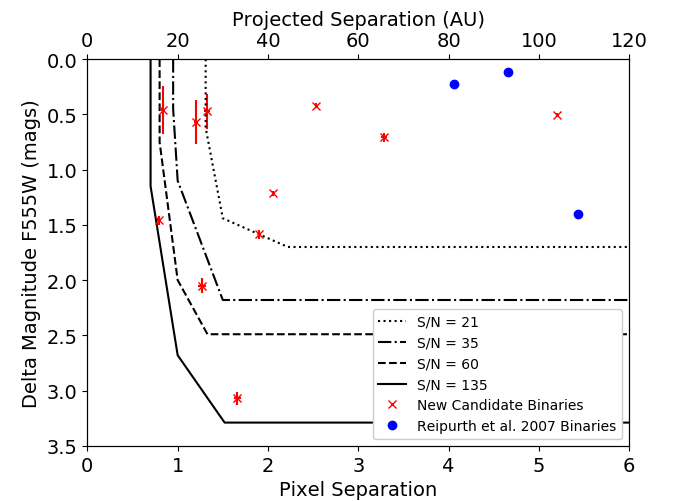}
\caption{Example contrast curves are plotted according to the legend.  Candidate binaries from the sample of Orion members are displayed with red X's.  Binaries found in \citet{Reipurth2007} and our survey are displayed with blue circles.  \label{fig:cc_curves}}
\end{figure}

Our sample consists of 101 ONC members that meet the selection criteria defined in Sec. 2.3 and are all classified as M-type stars.  After fitting our model to each source, 14 had companions that were found within the recovery limits shown in Fig. \ref{fig:Cmap_SN} with a $\chi^{2}_{\nu}$ $< $ 1.774.  Of these 14, 3 were previously detected by \citet{Reipurth2007} while 11 are new detections.  See Fig. \ref{fig:ex_binaries} for images of the candidate binaries.  Fig. \ref{fig:cc_curves} displays the output parameters of the candidate binaries compared to four example contrast curves of different S/N.  In Table \ref{table1}, we show the derived masses, mass ratios, projected separations, position angle, and output difference in magnitude (F555W filter) of each detected binary.  The projected separation was calculated assuming a distance of 400 pc to each source, determined using GAIA data by \citet{oriondistance}.  We cannot calculate distances to individual binaries using GAIA parallax measurements because we do not know the orientation of the orbit of the companions which can have a substantial effect on the parallax signal.  The mass ratio is defined as the mass of the secondary divided by the mass of the primary.  To estimate masses, we used the ages and extinctions of each source \citep{DaRio2016} and the isochrones defined in \citet{Bell2014} to determine masses (derived from BT-Settl atmospheres, Allard et. al 2011), assuming the extinction and age of the companion is equal to that of the primary.  In addition, we do not account for the possibility of excess accretion luminosity impacting the flux of either component.

\subsection{Completeness Limits} \label{subsec:completenesslimits}
As seen in Fig. \ref{fig:Cmap_SN}, our method is effective in recovering similar-brightness companions at separations down to 0.7 pixels (below the diffraction limit of HST) for a source with S/N = 135 and down to $\sim$ 1.9 pixels for a source with S/N = 13.  The completeness maps follow the expected trend that our algorithm achieves greater contrast around higher S/N sources.  For the highest S/N primaries (135), our method can recover companions with a difference in magnitude of $\sim$ 3.25 mags beginning at a separation of 1.5 pixels.  For primaries with S/N = 21, our method can recover companions with a difference in magnitude of $\sim$ 1.7 mags beginning at a separation of 2 pixels.

\subsection{False Positives} \label{subsec:falstpositives}
While the completeness map tests the region over which our algorithm will recover binaries, we must also take into account the possibility that our algorithm will recover false positives corresponding to a binary fit to a single star that returns a good fit.  If these were found to be common in regions of high completeness, then it would be crucial to calculate the probability that a recovered binary was a false positive.  

To explore this, we created 2400 single stars for each S/N from the empirical PSF models and added Poisson noise to mimic the background environment.  We then ran our algorithm and analyzed the output parameters of the fits to these singles.  In Fig. \ref{fig:Cmap_SN}, we plot the binary fit outputs to these single stars (labeled 'False Positives').  All of the false positives fall significantly outside of the 90\% completeness level, regardless of S/N, and are not confused with detected binaries.

\subsection{Chance Alignment Binaries} \label{subsec:chancealignments}
We must also calculate the expected number of companions that are not true (bound) companions, but chance alignments from field stars or other ONC members.  We exclude background stars because there is a wall of high extinction behind the ONC that blocks the vast majority of possible background contamination \citep{HillenbrandHartmann}.

To calculate the expected number of foreground contaminants over our field of view, we calculate the volume of space between us and an individual source and multiply by the number density of field stars near the Sun.  We limit our search to companions at a separation of 160 AU from the primary star.  The volume of space between us and a single source in our sample that a foreground contaminant could occupy is a cone with radius 160 AU and height 400 pc (distance to the ONC), equal to 2.5 x 10$^{-4}$ pc$^{3}$.  From \citet{Kounkel2018} and \citet{Kipper}, the stellar number density in the solar neighborhood is $\sim$ 0.3 stars/pc$^{3}$.  Multiplying the stellar number density by the space volume to a single source and then multiplying by the number of sources in our survey, the total expected number of chance alignments as a result of the solar neighborhood is 7.6 x 10$^{-5}$ stars per source.

We must also calculate the number of chance alignments due to members in the ONC.  \citet{Reipurth2007} defined the stellar surface density in the ONC as a function of distance from $\theta^{1}$ Ori C, an exponentially decreasing function with distance.  If we assume the surface density of the entire region is equal to that of the distance to the binary detected closest to $\theta^{1}$ Ori C, i.e. the worst case scenario, the stellar surface density is 11 stars/arcmin$^{2}$.  We search for companions within a circle of radius 160 AU, area of 8.0 x 10$^{4}$AU$^{2}$.  Assuming a distance of 400 pc to the ONC, we expect the number of contaminants per source from the ONC to be 0.0015 stars.  Therefore, we expect that none of our detected companions are line of sight companions, and that they are probably bound companions.

\begin{deluxetable*}{cccccccccccc}
\tablenum{1}
\tablecaption{Candidate binaries with masses (\(\textup{M}_\odot\)), mass ratios (q), projected separations, position angles in degrees, and difference in magnitude (F555W filter) between primary and secondary.  To estimate masses, we used the ages and extinctions of each source \citep{DaRio2016} and the isochrones defined in \citet{Bell2014} to determine masses, derived from BT-Settl atmospheres of \citet{Allard2011}.  We assumed a distance of 400 pc to calculate projected separations \citep{oriondistance}.  Sources \#11, 12, and 14 were previously identified in \citet{Reipurth2007} \label{table1}}

\tablewidth{0pt}
\tablehead{
\colhead{\#} & \colhead{2MASS ID} & \colhead{M$_{prim}$} & \colhead{M$_{sec}$} & \colhead{q} & 
\multicolumn2c{Projected Separation} & \colhead{PA (deg)} & \colhead{Total Mag} & \colhead{$\Delta$ mag} & \colhead{S/N} & \colhead{$\chi^{2}_{\nu}$} \\
\colhead{} & \colhead{} & \nocolhead{Name} & \colhead{} & \colhead{} & \colhead{(AU)} & \colhead{(Arcsec)} & \colhead{(E of N)} & \colhead{(F555W)} & \colhead{(F555W)} & \colhead{} & \colhead{} }
\decimalcolnumbers
\startdata
  1 & J05353280-0517384 &  0.38 $\pm$ .005 &  0.22 $\pm$ .004 &  0.57 &  16 $\pm$ 0.5 &  0.040 &  145 $\pm$ 1.7 &  19.593 &  1.46 $\pm$ 0.04 &  138  & 1.39\\
  2 & J05350309-0522378 &  0.22 $\pm$ .014 &  0.19 $\pm$ .016 &  0.86 &  17 $\pm$ 0.4 &  0.043 &  13 $\pm$ 1.4 &  19.233 &  0.46 $\pm$ 0.22 &  131  & 1.20\\
  3 & J05351986-0531038 &  0.40 $\pm$ .020 &  0.35 $\pm$ .016 &  0.88 &  24 $\pm$ 0.5 &  0.060 &  258 $\pm$ 0.6 &  20.788 &  0.57 $\pm$ 0.20 &  100  & 1.02\\
  4 & J05351475-0534167 &  0.48 $\pm$ .007 &  0.32 $\pm$ .005 &  0.66 &  25 $\pm$ 0.8 &  0.063 &  239 $\pm$ 0.7 &  20.512 &  2.05 $\pm$ 0.07 &  122  & 1.00\\
  5 & J05350121-0521444 &  0.10 $\pm$ .010 &  0.08 $\pm$ .008 &  0.78 &  27 $\pm$ 0.7 &  0.068 &  112 $\pm$ 1.9 &  21.610 &  0.47 $\pm$ 0.16 &  26  & 0.96\\
  6 & J05344791-0535438 &  0.31 $\pm$ .007 &  0.07 $\pm$ .003 &  0.23 &  33 $\pm$ 0.7 &  0.083 &  84 $\pm$ 0.9 &  19.887 &  3.07 $\pm$ 0.06 &  152  & 1.75\\
  7 & J05351227-0520452 &  0.34 $\pm$ .003 &  0.25 $\pm$ .002 &  0.73 &  38 $\pm$ 0.4 &  0.095 &  117 $\pm$ 0.4 &  19.127 &  1.58 $\pm$ 0.04 &  94  & 1.42\\
  8 & J05352190-0515011 &  0.26 $\pm$ .001 &  0.18 $\pm$ .001 &  0.70 &  41 $\pm$ 0.3 &  0.100 &  68 $\pm$ 0.3 &  20.651 &  1.21 $\pm$ 0.02 &  102  & 1.10\\
  9 & J05350207-0518226 &  0.25 $\pm$ .001 &  0.22 $\pm$ .001 &  0.89 &  51 $\pm$ 0.2 &  0.130 &  24 $\pm$ 0.2 &  20.109 &  0.42 $\pm$ 0.02 &  106  & 1.14\\
  10 & J05351624-0528337 &  0.29 $\pm$ .003 &  0.25 $\pm$ .003 &  0.87 &  66 $\pm$ 0.6 &  0.165 &  184 $\pm$ 0.5 &  22.115 &  0.71 $\pm$ 0.04 &  21  & 0.90\\
  11 & J05345483-0525125 &  0.15 $\pm$ .001 &  0.13 $\pm$ .001 &  0.89 &  81 $\pm$ 0.2 &  0.203 &  330 $\pm$ 0.1 &  19.449 &  0.22 $\pm$ 0.01 &  161 &  1.24\\
  12 & J05351270-0527106 &  0.20 $\pm$ .001 &  0.19 $\pm$ .001 &  0.95 &  93 $\pm$ 0.2 &  0.233 &  359 $\pm$ 0.1 &  19.275 &  0.12 $\pm$ 0.01 &  154  & 1.20\\
  13 & J05353650-0520094 &  0.28 $\pm$ .001 &  0.25 $\pm$ .001 &  0.91 &  104 $\pm$ 0.3 &  0.260 &  335 $\pm$ 0.1 &  21.925 &  0.50 $\pm$ 0.01 &  49  & 1.16\\
  14 & J05351676-0517167 &  0.14 $\pm$ .001 &  0.07 $\pm$ .001 &  0.51 &  109 $\pm$ 0.4 &  0.273 &  86 $\pm$ 0.1 &  21.253 &  1.40 $\pm$ 0.02 &  53  & 0.88
\enddata
\end{deluxetable*}

\subsection{Comparison to the Galactic Field} \label{subsec:fieldcomparison}
In order to compare our results to that of the Galactic Field, we must identify a sub-sample of our 101 sources for which we can detect companions over a common mass ratio and separation range.  We calculate the maximum magnitude difference at 1.5 pixel separation for all sources in order to attain both high contrast and retain a majority of our original 101 source sample.  From each magnitude difference, we used the ages of each source \citep{DaRio2016} and the isochrones defined in \citet{Bell2014} as described in Sec. 2.3 to determine minimum masses (derived from BT-Settl atmospheres, Allard et al. 2011) of the companion detectable for each source.  The minimum mass ratio sensitivity is then defined as the minimum detectable companion mass divided by the mass of the primary.  We require every source to be sensitive to companions that result in a mass ratio of at least 0.6 at a separation of 1.5 pixels (30 AU).  This cut results in a sample of 75 sources, 6 of which had companions between mass ratios of 0.6 - 1 and projected separations 30 - 160 AU (1.5 - 8 pixels), resulting in a companion frequency of 8$_{-2\%}^{+4\%}$.  Errors on the companion frequency are estimated from Binomial statistics where we define the upper and lower bounds based on the 68\% confidence range of the Binomial distribution \citep[of][]{Burg}.

We utilized surveys of M-dwarfs in the Galactic Field to define the known binary population and predict an expected outcome of a survey sensitive over the same separation range and mass ratio range as ours.  

First, we define the mass ratio (q) distribution using the functional form from \citet{ReggianiMeyer2013}:

\begin{equation}
    \frac{dN_{1}}{dq} \propto q^{\beta}
    \label{qdistribution}
\end{equation}
where $\beta$ = 0.25 $\pm$ 0.29, appropriate for M-type stars.

Then, we define the surface density distribution as a log-normal distribution with the observed mean and sigma values from \citet{Winters2019} ($\overline{a}$ = 20 AU, $\sigma_{loga}$ = 1.3) for M-type stars, where a is the projected separation of the binary:

\begin{equation}
\frac{dN_{2}}{da} = \frac{1}{a} e^{-\frac{log(a)-log(\overline{a})}{2\sigma^{2}}}
    \label{adistribution}
\end{equation}

Additionally, we have to normalize the product of the integrals of these functions (Eq. \ref{cf}) based on the companion frequency results of an M-star survey.  The \citet{Janson2012} results gave a companion frequency (CF) of 17\% $\pm$ 3\% over q = 0.6 - 1 and a = 3 - 227 AU.  We use those values to determine the coefficient of integration, C$_{n}$, for the following relation:

\begin{equation}
    CF = C_{n}*\int_{q_1}^{q_2} \frac{dN_{1}}{dq} \int_{a_1}^{a_2} \frac{dN_{2}}{da}
    \label{cf}
\end{equation}

These functional forms are integrated over q = 0.6 - 1 and a = 30 - 160 AU to return the expected companion frequency.  This yields an expected companion frequency of the Galactic Field of 6.5\% $\pm$ 3\% \footnote[1]{After integrating the functional form over 1$\sigma$ differences in $\beta$ and the normalization companion frequency, the expected Galactic Field companion frequency changes by only 1\%.  Thus, the ONC binary population is still consistent with the Galactic Field with a difference of \textless 1$\sigma$.}.  We conclude that the binary star population of the ONC, 8$_{-2\%}^{+4\%}$, is consistent with that of the Galactic Field for low-mass stars 0.1 - 0.6 \(\textup{M}_\odot\) over mass ratios 0.6 - 1 and separations 30 - 160 AU.

\section{Discussion} \label{sec:discussion}
\subsection{Comparison to Taurus} \label{subsec:taurus}
Previous studies of low-mass star forming regions, like Taurus-Auriga, have shown an excess of companion systems compared to the Galactic Field population.  After completeness corrections provided by Bayesian analysis, \citet{Kraus2011} find a companion frequency of 79$_{-11\%}^{+12\%}$ over separations 3 - 5000 AU and over all mass ratios (q = 0 - 1) for their low-mass star sample of mass 0.25 - 0.7 \(\textup{M}_\odot\).  In order to compare this survey to ours, we restricted the Taurus parent population and our ONC parent population to primaries with  0.25 \(\textup{M}_\odot\) $< $ \textup{M} $< $ 0.6 \(\textup{M}_\odot\) (their minimum mass and our maximum mass, respectively).  Then, we consider only binaries with q = 0.6 - 1 and separation = 30 - 160 AU.  This process left us with 74 members and 11 binaries in Taurus and 46 members and 3 binaries in the ONC.  Therefore, over our sensitivity range, we calculate a companion frequency in Taurus of 15$_{-3\%}^{+5\%}$ and in our sample of the ONC of 6.5$_{-2\%}^{+6\%}$, a difference of only 1.3$\sigma$.  A larger sample or expanding the parameter space in the survey could demonstrate a statistically significant difference between the binary populations of Taurus and the ONC.  As stated in Sec. 3.4, the expected companion frequency of the Galactic Field over the same range of parameters is 6.5\% $\pm$ 3\%.  This reveals a higher companion frequency in Taurus compared to the Galactic Field, at the 2.0$\sigma$ level.  \citet{Kraus2011} found a more significant binary excess (3$\sigma$ compared to our Galactic Field model) as they had more sources and covered a larger parameter space in q and separation. 

\subsection{Comparison to Previous ONC Surveys} \label{subsec:oncsurveys}
Previous surveys have studied the binary population of the ONC.  \citet{Reipurth2007} assessed the binary population of the ONC using ACS during HST Proposal 9825 (PI: J. Bally).  Their survey achieved separations down to a minimum of 67.5 AU, using the narrow band H$_{\alpha}$ filter to detect companions.  Determining mass ratios from H$_{\alpha}$ flux ratios is difficult since young stars typically emit in H$_{\alpha}$ due to accretion from the circumstellar disk, chromospheric emission, and often suffer contamination from highly structured nebular background.

\citet{Kounkel2016_100} carried out a direct imaging survey of the ONC using HST.  They observed both protostars and pre-main-sequence stars with NICMOS and WFC3, achieving separations of 100 - 1000 AU (0.25 - 2.5" for the ONC).  They state that the masses of both the primaries and companions have not been determined, and thus we cannot directly compare their survey to ours.

A recent spectroscopic survey in the ONC by \citet{Kounkel2016_Spec} uses radial velocities to find tight binaries ($< $ 10 AU).  They are $\sim$ 20\% complete out to 4 AU and 5\% complete out to 10 AU.  They calculate mass ratios for two of their binaries which are the only double-lined spectroscopic binaries in the survey.  This survey predominantly deals with sources that have temperatures $> $ 4000 K which are more massive than our sources and therefore not comparable.

\citet{Kohler2006} also studied the binary population of the ONC using AO on the Keck II telescope and ESO 3.6m telescope on La Silla. Similar to \citet{Reipurth2007}, they were sensitive to binary separations down to a minimum of 60 AU.  They found a companion frequency of 3.8\% $\pm$ 1.5\% for their low-mass star sample. However, their sample was not very sensitive to companions around stars with mass $< $ 2\(\textup{M}_\odot\) and appear to achieve a magnitude difference $\Delta$K $\sim$ 1 at 0.4".  Assuming a median age of the cluster of 2 Myr and a primary mass equal to the median mass of their low-mass star sample of 0.3 \(\textup{M}_\odot\) (as quoted in their paper), the sensitivity of their survey achieves q down to $\sim$ 0.4 for separations 165 - 500 AU.  Using the same calculation stated in Eq. \ref{cf}, the companion frequency of the ONC over their sensitivity is consistent with the expected Galactic Field companion frequency of 4\% $\pm$ 3\%.

\citet{Duchene2018} surveyed the ONC using ground-based AO-assisted imaging with NACO on VLT/UT4.  They had a sample of 42 sources, most of which (31) would have been excluded from our sample because they are $> $ 0.6 \(\textup{M}_\odot\).  They quote a minimum companion frequency of 22\% as they are not complete for all mass ratios.  They conclude that the ONC has an excess of companions compared to the Galactic Field, even for low-mass stars.  We can approximate an expected M-dwarf companion frequency of the Galactic Field population based on their separation sensitivity (10 - 60 AU) and their quoted magnitude sensitivity ($\Delta$K $> $ 2 mags).  Assuming a median age of the cluster of 2 Myr, the sensitivity of their survey achieves at least a limiting q = 0.2 for primary masses of 0.1-0.6 \(\textup{M}_\odot\).  Using the same calculation stated in Eq. \ref{cf}, the expected Galactic Field companion frequency is 14\% $\pm$ 3\%, over their sensitivities for primary mass stars $< $ 0.6 \(\textup{M}_\odot\).  Their sample contains 11 sources with primary masses $< $ 0.6 \(\textup{M}_\odot\), with 3 identified as candidate binaries.  This results in a companion frequency of 27$_{-9\%}^{+16\%}$, higher than the Galactic Field, but only a 1.4$\sigma$ difference.

It is clear that despite all the work that has been devoted to studying the binary population of the ONC, a complete description (e.g. companion frequency and mass ratio distribution) of the binary population has yet to be established.  This is true over the entire parameter space of primary mass, mass ratio, and separation.  Much larger samples over a broad range of these parameter spaces are necessary to determine a precise description of the binary population of the ONC for comparison to the Galactic Field.

\subsection{Implications} \label{subsec:implications}

Our analysis of the ONC results in a companion frequency that is consistent with that of the Galactic Field M-dwarf population.  It has also been shown that the companion frequency of M-stars in Taurus is in excess of that of the Galactic Field population \citep{Kraus2011}.  One of the hypotheses presented to explain the excess of companions in Taurus compared to the Galactic Field is dynamical evolution destroying wide binaries in stellar associations before they dissociate into the Galactic Field, e.g. \citet{MarksKroupa2012}.  However, \citet{Kraus2011} show that the binary excess for low-mass primaries in Taurus compared to the Galactic Field exists at separations $\lesssim$ 200 AU.  This would mean that closer separation binaries would have to be preferentially destroyed through dynamical interactions in order to decrease the companion frequency.  Tighter binaries have larger binding energies and are therefore harder to destroy dynamically.  The excess of companions to low-mass primaries in Taurus is difficult to explain through dynamical evolution, especially since Taurus has a low stellar density.

\citet{ParkerMeyer2014} offer a different vision for how binary populations evolve.  \citet{ParkerMeyer2014} performed N-body simulations to determine the effect of dynamical evolution on binary populations based on primary star mass.  They set up two simulations, one where the initial companion frequency was unity with an excess, compared to the Galactic Field, of wide binaries (up to 10,000 AU) and another where the initial companion frequency and binary population parameters were derived from Galactic Field binary population studies based on primary mass.  The first test in relatively high stellar density regions did not reproduce the companion frequency found in the Galactic Field.  However, the second test showed that the companion frequency for low-mass stars was changed by only a few percent even after allowing for dynamical evolution over 10 Myr, a concept supported by \citet{ParkerGoodwin2011} and \citet{ParkerGoodwin2012}.  They conclude that if the initial companion frequency for low-mass stars in dense star clusters ( $\sim$ 10$^{3}$ \(\textup{M}_\odot\)/pc$^{3}$) is the same as observed in the Galactic Field, then dynamical evolution does not destroy too many binaries (especially with separations $\lesssim$ 100s AU).  Perhaps binary system properties are largely set by core fragmentation \citep{Offner2010} and only marginally affected by later dynamical evolution in a cluster.

\citet{Simon1997} find that more wide binaries ($\gtrsim$ 1000 AU) exist in low stellar density star forming regions like Taurus compared to high stellar density regions like the Orion Trapezium.  This wide binary excess in low stellar density regions can result from less dynamical processing, whereas higher stellar density regions like the ONC contain less wide binaries due to enhanced dynamical processing.  Even denser regions still, like Westerlund 1 (10$^{5}$ \(\textup{M}_\odot\)/pc$^{3}$) and Arches (10$^{5.5}$ \(\textup{M}_\odot\)/pc$^{3}$) \citep{Clark}, would have significantly fewer wide binaries and could destroy many binaries with separations $\lesssim$ 100s AU.  It is possible that these different stellar density regions all contribute a significant portion of the binaries that exist in the Galactic Field \citep{ladalada}, although the types of binaries these star forming regions preferentially impart to the Galactic Field will depend on stellar density.  Based on our results, the ONC would represent a more typical population comparable to that which exists in the Galactic Field than perhaps low density regions like Taurus.

\section{Conclusion} \label{sec:conclusion}
We carried out a binary survey of the Orion Nebula Cluster using an automated double-PSF fitting technique.  Our combined use of archival HST data and the publicly available PSF models of \citet{AndersonKing2006} provide for an efficient way to cover large survey areas at high spatial resolution.  In summary:

\vspace{0.35cm}
1) We have shown that the empirical PSF models of AK06 can be effectively used to identify binaries at angular resolution less than the diffraction limit of HST for the F555W filter, 0.06", for higher S/N sources, on par with recent adaptive optics surveys, e.g. \citet{Duchene2018}, and 4x higher resolution than previous HST binary surveys, e.g. \citet{Reipurth2007}.  Our method provides a complementary approach to ground-based AO studies for discovering companions, while also covering a large survey area.

\vspace{0.35cm}
2) We detected 14 candidate binaries from our total sample of 101 M-stars, 11 new detections and 3 previously detected.  The closest binary identified has a separation of 16 AU.

\vspace{0.35cm}
3)  Over the mass ratio range of 0.6 - 1 and separation range of 30 - 160 AU, we observed a companion frequency of 8$_{-2\%}^{+4\%}$ in the ONC. This result is consistent with the expected companion frequency of M-dwarfs in the Galactic Field over those same sensitivity ranges, 6.5\% $\pm$ 3\%, a 0.4$\sigma$ difference.

\vspace{0.35cm}
4) High-mass, high stellar density star forming regions like the ONC do not require further dynamical evolution to shape binary populations into one resembling the Galactic Field.  This suggests the ONC may be more typical of the Galactic Field population than low density regions like Taurus.

\acknowledgments

We would like to thank Jay Anderson for many productive discussions on PSF modeling and the implementation of his PSF code, as well as Megan Kiminki for contributions to the construction of our code.  We thank an anonymous referee for a careful and helpful report which improved the presentation of this manuscript.  Megan Reiter received funding from the European Union’s Horizon 2020 research and innovation programme under the Marie Skĺodoska-Curie grant agreement No. 665593 awarded to the Science and Technology Facilities Council. Trent J. Dupuy acknowledges research support from Gemini Observatory.  This work is based on observations made with the NASA/ESA Hubble Space Telescope, obtained from the data archive at the Space Telescope Science Institute. STScI is operated by the Association of Universities for Research in Astronomy, Inc. under NASA contract NAS 5-26555. Support for this work was provided by NASA through grant number HST-AR-15047.001-A from the Space Telescope Science Institute, which is operated by AURA, Inc., under NASA contract NAS 5-26555.

\appendix
\section{Singles}

\begin{deluxetable}{cccccc}[b]
\tablenum{2}
\tablecaption{This table contains the sources in our sample for which we did not find a companion.  Of our 101 sources, 87 did not have companions and are labeled as single sources over the sensitivity ranges in which our code could identify a companion.  The S/N of each source is also displayed which gives an approximation of the separation and mass range in which a companion does not exist.  Magnitudes were determined by \citet{DaRio2016}}
\tablewidth{0pt}
\tablehead{
\colhead{2MASS ID} & \colhead{RA (J2000)}  & \colhead{Dec (J2000)}  & \colhead{J (mag)} & \colhead{H (mag)}  & \colhead{S/N} }
\decimalcolnumbers
\startdata
J05351794-0525061	&	5:35:17.95	&	-5:25:06.17	&	12.024	&	12.100	&	11\\
J05351015-0527574	&	5:35:10.16	&	-5:27:57.44	&	11.044	&	12.367	&	28\\
J05351986-0530321	&	5:35:19.87	&	-5:30:32.17	&	12.973	&	12.287	&	154\\
J05351047-0526003	&	5:35:10.48	&	-5:26:00.30	&	13.059	&	12.243	&	58\\
J05352445-0524010	&	5:35:24.45	&	-5:24:01.02	&	12.825	&	12.283	&	9\\
J05353437-0526596	&	5:35:34.38	&	-5:26:59.70	&	13.034	&	12.328	&	113\\
J05351826-0529538	&	5:35:18.26	&	-5:29:53.83	&	12.924	&	12.274	&	163\\
J05352433-0526003	&	5:35:24.34	&	-5:26:00.34	&	13.130	&	12.434	&	102\\
J05352534-0525295	&	5:35:25.35	&	-5:25:29.58	&	13.194	&	12.420	&	112
\enddata
\end{deluxetable}

\begin{deluxetable*}{cccccc}[t]
\tablenum{2}
\tablecaption{}
\tablewidth{0pt}
\tablehead{
\colhead{2MASS ID} & \colhead{RA (J2000)}  & \colhead{Dec (J2000)}  & \colhead{J (mag)} & \colhead{H (mag)}  & \colhead{S/N}}
\decimalcolnumbers
\startdata
J05345918-0523078	&	5:34:59.19	&	-5:23:07.85	&	13.003	&	12.324	&	136\\
J05352266-0515085	&	5:35:22.67	&	-5:15:08.59	&	13.365	&	12.422	&	73\\
J05345853-0532498	&	5:34:58.53	&	-5:32:49.86	&	13.300	&	12.404	&	124\\
J05351379-0519254	&	5:35:13.79	&	-5:19:25.46	&	13.138	&	12.410	&	122\\
J05351216-0530201	&	5:35:12.16	&	-5:30:20.16	&	13.180	&	12.379	&	158\\
J05345009-0517121	&	5:34:50.09	&	-5:17:12.19	&	13.329	&	12.371	&	81\\
J05345931-0523326	&	5:34:59.31	&	-5:23:32.68	&	12.988	&	12.211	&	73\\
J05351797-0526506	&	5:35:17.97	&	-5:26:50.67	&	13.306	&	12.374	&	66\\
J05352228-0531168	&	5:35:22.28	&	-5:31:16.87	&	13.165	&	12.423	&	159\\
J05350270-0532249	&	5:35:02.70	&	-5:32:24.93	&	13.444	&	12.396	&	82\\
J05353175-0516399	&	5:35:31.75	&	-5:16:39.90	&	13.290	&	12.237	&	55\\
J05352705-0515447	&	5:35:27.06	&	-5:15:44.75	&	12.747	&	11.763	&	143\\
J05351083-0525569	&	5:35:10.84	&	-5:25:56.99	&	13.505	&	12.393	&	57\\
J05350803-0536140	&	5:35:08.04	&	-5:36:14.08	&	12.703	&	11.792	&	157\\
J05352512-0522252	&	5:35:25.13	&	-5:22:25.27	&	12.998	&	12.023	&	19\\
J05345583-0519454	&	5:34:55.83	&	-5:19:45.49	&	13.603	&	12.363	&	49\\
J05350859-0526194	&	5:35:08.60	&	-5:26:19.45	&	13.240	&	12.317	&	77\\
J05353156-0516369	&	5:35:31.56	&	-5:16:36.95	&	12.645	&	11.615	&	125\\
J05352657-0517530	&	5:35:26.57	&	-5:17:53.08	&	12.911	&	11.790	&	71\\
J05350201-0518341	&	5:35:02.01	&	-5:18:34.18	&	13.136	&	12.070	&	93\\
J05350370-0522457	&	5:35:03.70	&	-5:22:45.71	&	13.622	&	12.267	&	16\\
J05352753-0513563	&	5:35:27.53	&	-5:13:56.33	&	13.806	&	12.402	&	24\\
J05352002-0529119	&	5:35:20.02	&	-5:29:11.90	&	13.687	&	12.307	&	31\\
J05351277-0520349	&	5:35:12.78	&	-5:20:34.94	&	13.585	&	12.320	&	47\\
J05353014-0514185	&	5:35:30.15	&	-5:14:18.57	&	13.083	&	11.854	&	55\\
J05350822-0524032	&	5:35:08.23	&	-5:24:03.24	&	13.313	&	12.302	&	36\\
J05352696-0524005	&	5:35:26.97	&	-5:24:00.59	&	12.983	&	12.130	&	74\\
J05351884-0522229	&	5:35:18.84	&	-5:22:22.96	&	12.121	&	11.640	&	7\\
J05353264-0515514	&	5:35:32.64	&	-5:15:51.45	&	13.576	&	12.399	&	47\\
J05351029-0519563	&	5:35:10.30	&	-5:19:56.30	&	12.965	&	11.882	&	105\\
J05352103-0522250	&	5:35:21.03	&	-5:22:25.03	&	12.560	&	11.939	&	19\\
J05351041-0519523	&	5:35:10.41	&	-5:19:52.36	&	12.912	&	11.728	&	64\\
J05353283-0518198	&	5:35:32.83	&	-5:18:19.83	&	13.964	&	12.417	&	10\\
J05352321-0521357	&	5:35:23.21	&	-5:21:35.78	&	13.171	&	12.204	&	63\\
J05350732-0538409	&	5:35:07.33	&	-5:38:40.98	&	14.041	&	12.439	&	27\\
J05351375-0534548	&	5:35:13.76	&	-5:34:54.90	&	12.798	&	11.540	&	153\\
J05352246-0525451	&	5:35:22.46	&	-5:25:45.11	&	12.781	&	11.906	&	47\\
J05350560-0518248	&	5:35:05.61	&	-5:18:24.85	&	12.909	&	11.684	&	108\\
J05353162-0516581	&	5:35:31.62	&	-5:16:58.20	&	12.981	&	11.680	&	132\\
J05352835-0517544	&	5:35:28.35	&	-5:17:54.49	&	13.604	&	12.313	&	81\\
J05353570-0520331	&	5:35:35.71	&	-5:20:33.18	&	13.731	&	12.396	&	26\\
J05350434-0538311	&	5:35:04.34	&	-5:38:31.12	&	13.554	&	12.260	&	107\\
J05353244-0515068	&	5:35:32.44	&	-5:15:06.90	&	13.394	&	12.108	&	18\\
J05351587-0522328	&	5:35:15.87	&	-5:22:32.83	&	13.127	&	11.749	&	6
\enddata
\end{deluxetable*}

\begin{deluxetable*}{cccccc}[t]
\tablenum{2}
\tablecaption{}
\tablewidth{0pt}
\tablehead{
\colhead{2MASS ID} & \colhead{RA (J2000)}  & \colhead{Dec (J2000)}  & \colhead{J (mag)} & \colhead{H (mag)}  & \colhead{S/N}}
\decimalcolnumbers
\startdata
J05350487-0520574	&	5:35:04.87	&	-5:20:57.41	&	13.340	&	12.034	&	59\\
J05351269-0519353	&	5:35:12.69	&	-5:19:35.39	&	13.470	&	12.105	&	53\\
J05351185-0517259	&	5:35:11.85	&	-5:17:26.00	&	13.552	&	12.336	&	78\\
J05351851-0520427	&	5:35:18.52	&	-5:20:42.77	&	12.763	&	11.526	&	121\\
J05345693-0522062	&	5:34:56.93	&	-5:22:06.28	&	13.457	&	12.333	&	102\\
J05350416-0520156	&	5:35:04.16	&	-5:20:15.66	&	13.504	&	12.241	&	47\\
J05351921-0531030	&	5:35:19.21	&	-5:31:03.02	&	13.934	&	12.353	&	58\\
J05351305-0521532	&	5:35:13.06	&	-5:21:53.29	&	13.634	&	12.196	&	9\\
J05350609-0514249	&	5:35:06.10	&	-5:14:24.98	&	13.292	&	12.039	&	116\\
J05350540-0524150	&	5:35:05.41	&	-5:24:15.05	&	13.178	&	11.929	&	73\\
J05352615-0522570	&	5:35:26.16	&	-5:22:57.03	&	13.141	&	11.599	&	14\\
J05351294-0528498	&	5:35:12.94	&	-5:28:49.90	&	13.463	&	12.407	&	138\\
J05351053-0522166	&	5:35:10.53	&	-5:22:16.64	&	13.211	&	11.580	&	1\\
J05352465-0522425	&	5:35:24.66	&	-5:22:42.57	&	13.418	&	11.635	&	25\\
J05352268-0516140	&	5:35:22.69	&	-5:16:14.01	&	13.200	&	11.995	&	127\\
J05350617-0522124	&	5:35:06.18	&	-5:22:12.42	&	13.293	&	11.867	&	13\\
J05345826-0538257	&	5:34:58.26	&	-5:38:25.72	&	13.793	&	12.234	&	31\\
J05352640-0516124	&	5:35:26.41	&	-5:16:12.45	&	13.417	&	11.945	&	45\\
J05351689-0517029	&	5:35:16.90	&	-5:17:02.94	&	13.361	&	11.635	&	28\\
J05350161-0533380	&	5:35:01.62	&	-5:33:38.04	&	13.075	&	11.588	&	120\\
J05345737-0514334	&	5:34:57.38	&	-5:14:33.49	&	13.423	&	12.239	&	120\\
J05345837-0521166	&	5:34:58.38	&	-5:21:16.64	&	13.202	&	11.975	&	126\\
J05351343-0521073	&	5:35:13.43	&	-5:21:07.33	&	13.641	&	12.009	&	8\\
J05351797-0516451	&	5:35:17.98	&	-5:16:45.11	&	13.150	&	11.650	&	111\\
J05351197-0522541	&	5:35:11.97	&	-5:22:54.12	&	11.489	&	10.684	&	9\\
J05344677-0526048	&	5:34:46.78	&	-5:26:04.87	&	13.725	&	12.379	&	147\\
J05351389-0518531	&	5:35:13.90	&	-5:18:53.18	&	13.517	&	12.053	&	73\\
J05350332-0516227	&	5:35:03.33	&	-5:16:22.78	&	14.097	&	12.376	&	21\\
J05353074-0521466	&	5:35:30.75	&	-5:21:46.62	&	13.978	&	12.347	&	36\\
J05352317-0522283	&	5:35:23.17	&	-5:22:28.33	&	14.465	&	12.132	&	8\\
J05351421-0520042	&	5:35:14.22	&	-5:20:04.28	&	13.524	&	11.941	&	116\\
J05351567-0517472	&	5:35:15.67	&	-5:17:47.28	&	12.719	&	10.930	&	26\\
J05353099-0522013	&	5:35:31.00	&	-5:22:01.31	&	13.358	&	11.440	&	61\\
J05350873-0522566	&	5:35:08.73	&	-5:22:56.69	&	12.570	&	11.191	&	36
\enddata
\end{deluxetable*}

\bibliography{sample63}{}
\bibliographystyle{aasjournal}

\end{document}